\documentclass[prb,preprintnumbers,superscriptaddress,aps,showpacs]{revtex4}
\usepackage{amsmath,amsbsy,amssymb,array}

\begin{document}

\title{Superbosonization formula and its application to random matrix theory}
\author{J.E.\ Bunder}
\affiliation{Physics Division, National Center for Theoretical
Sciences, Hsinchu 300, Taiwan}
\author{K.B.\ Efetov}
\affiliation{Theoretische Physik III, Ruhr-Universit\"at Bochum,
D-44780 Bochum, Germany}%
\affiliation{L.D. Landau Institute for Theoretical Physics, 117940
Moscow, Russia}
\author{V.E.\ Kravtsov}
\affiliation{The Abdus Salam ICTP, Strada Costiera 11, 34100,
Trieste, Italy}%
\affiliation{L.D. Landau Institute for Theoretical Physics, 117940
Moscow, Russia}
\author{O.M.\ Yevtushenko}
\affiliation{The Abdus Salam ICTP, Strada Costiera 11, 34100,
Trieste, Italy}
\author{M.R.\ Zirnbauer}
\affiliation{Institut fur Theoretische Physik, Universit\"at zu
K\"oln, 50937 K\"oln, Germany}
\date{May 15, 2007}

\begin{abstract}
Starting from Gaussian random matrix models we derive a new
supermatrix field theory model. In contrast to the conventional
non-linear sigma models, the new model is applicable for any range of
correlations of the elements of the random matrices. We clarify the
domain of integration for the supermatrices, and give a demonstration
of how the model works by calculating the density of states for an
ensemble of almost diagonal matrices. It is also shown how one can
reduce the supermatrix model to the conventional sigma model.
\end{abstract}

\maketitle

\section{Introduction}

The non-linear supersymmetric sigma model (NL$\sigma$M)
\cite{efetov83} has been very fruitful in describing quantum systems
with quenched disorder \cite{efetov}. Derived originally
\cite{efetov} from a model of noninteracting electrons in a Gaussian
random potential, NL$\sigma$M was later constructed in its
zero-dimensional version directly \cite{verbaarschot} from the
Wigner-Dyson ensembles \cite{mehta} of random matrix theory. This
gave the first demonstration of equivalence between these two
problems in the low-frequency domain. Furthermore, even for problems
where the conventional Wigner-Dyson random matrix theory is
sufficient (see, e.g.\ \cite{mehta,been,weidenmuller}) NL$\sigma$M
turned out to be a competitive method of calculation compared to the
standard approach based on orthogonal polynomials \cite{mehta}. Let
us mention parametric level statistics \cite{AltSim} as a beautiful
example of new results which were obtained in the field of classical
RMT using NL$\sigma$M.

The most significant applications of NL$\sigma$M, however, are in the
realm beyond classical RMT, such as corrections to the Wigner-Dyson
theory in systems with diffusive dynamics
\cite{KravtsovMirlin94,AndrAlt,Mirlin-review}, and random matrix
ensembles with a probability distribution not invariant under the
full group of unitary transformations. The domain of application of
such random matrix ensembles is rather wide. For example, band random
matrices are equivalent to thick disordered wires
\cite{efetov,fyodorov}, where strong localization was first
established using NL$\sigma$M. Another example is the Critical
Power-Law Band Random Matrix (CPLBRM) ensemble \cite{mirlin}. An
important property of eigenvector statistics in this ensemble is
multifractality \cite{mirlin,KravMut,Janssen}, which makes CPLBRM a
good model to qualitatively (and in many cases quantitatively)
describe the critical states near the Anderson transition
\cite{Janssen,evers}. In the analysis of all these models NL$\sigma$M
was an extremely useful guide, although limited as NL$\sigma$M is not
always exactly equivalent to the model being studied. The reason for
this limitation is the saddle-point approximation which is used in
deriving NL$\sigma$M. In the case of random band matrices the band
width $B$ has to be large in order for the saddle-point approximation
to be valid. For CPLBRM this implies weak multifractality, while for
random band matrices it means that the localization length is much
larger than the lattice scale. In both cases localization effects are
in a certain sense weak.

There are several important problems which lie beyond the
applicability of NL$\sigma$M. These include the properties of
one-dimensional disordered chains and the problem of critical
eigenfunctions and level statistics in systems with strong
multifractality. The Hamiltonian of the first example is just a
tridiagonal random matrix, while a representative model for the
second example is CPLBRM with a small bandwidth $B \sim 1\,$. Both of
these are examples of a class of Gaussian ensembles of {\it almost
diagonal random matrices} which were recently defined and studied in
Refs.\ \cite{KrYev2003, YevKr2004,KrCuYev2006,YevOs2007}. In order to
attack these problems by the supersymmetry method, one needs an exact
representation of the corresponding correlation functions in terms of
a supermatrix field theory.

Such a representation called 'superbosonization' has recently been
suggested in Refs. \cite{est,efetov04}. Similar ideas had been put
forward earlier by Lehmann, Saher, Sokolov, and Sommers, \cite{LSSS}
and by Hackenbroich and Weidenm\"uller. \cite{HW}. Here we give a
decent derivation of the key formula of super-bosonization, including
such aspects as integration measure and domain of integration. We
also work out a simple application of the formula involving Gaussian
random matrices.

Given any function $f(\psi \otimes \bar\psi)$ of the tensor product
of a super-vector $\psi$ and its conjugate $\bar\psi$, the
super-bosonization formula essentially converts the integral of $f$
over $\psi$ to an integral over a supermatrix $Q$ akin to $\psi
\otimes \bar\psi$. Thus it provides a direct way to rewrite in terms
of $Q$ the so-called $\psi$-functional which arises after ensemble
averaging the initial free-field representation of disordered
one-particle systems. We wish to stress that the conversion from
$\psi$ to $Q$ is \emph{exact}. It is also non-trivial, as the number
of independent variables in the supermatrix $Q$ may be smaller or
larger than the number of independent variables in the initial
super-vector representation. We note that a workable
\cite{OssipovKravtsov,YevOs2007} integration formula which converts
integrals over $\psi$ into integrals over some supermatrix
$\tilde{Q}$ (different from $Q$) can be constructed by a mere change
of variables \cite{OssipovKravtsov} without changing the number of
variables. What we describe in this paper is not simply a change of
variables.

For ease of discussion, we consider disordered one-particle
Hamiltonians in the form of a random Hermitian matrix $H$ with
independent, Gaussian-distributed entries $H_{ij}$ $(i,j = 1, \ldots,
N$) characterized by a certain variance matrix $C_{ij} = c(|i-j|) =
\langle |H_{ij}|^{2} \rangle$. The special case of classical
Wigner-Dyson random matrix theory is obtained by setting $c(|i-j|) =
\mathrm{const}$. While this has the largest possible symmetry group,
$\mathrm{U}(N)$, our ensembles with non-constant $c(|i-j|)$
generically have symmetry group $\mathrm{U}(1)^N$. From the
perspective of possible applications, the main outcome of this paper
is an exact field-theoretical representation of correlation functions
for the generic case of $\mathrm{U}(1)^N$ symmetry.

To exert better control over the mathematics, and also with a view to
certain applications, we consider random matrix models with a local
gauge symmetry $\mathrm{U}(n)^{N/n}$. The random matrix Hamiltonians
with matrix elements $H_{ij}^{ab}$ in such models have a block
structure with $i,j = 1, ..., M := N/n$ and $a,b = 1, ..., n\,$. The
probability distribution function is of the form
\begin{displaymath}
  P(H) = \exp\left( - {\textstyle{\frac{1}{2}}} \sum_{i,j=1}^{M}
  c(|i-j|)^{-1} \sum_{a,b=1}^{n}\,H_{ij}^{ab}\,H_{ji}^{ba}\right)\;.
\end{displaymath}
Setting $M=1$, $n=N$ gives the classical Gaussian Unitary Ensemble
(GUE) of Wigner and Dyson, while the case $n=1$, $M = N$ corresponds
to a generic Gaussian random matrix model.

It turns out that the precise form of the superbosonization formula
depends in a crucial way on how $n$ compares with the number of
commuting components $q$ in the supervector $\psi$. The formula
agrees with the naive expectation \cite{est,efetov04} if $q \leq n$
(as is the case, e.g., for the {\it mean} density of states (DoS) of
a generic Gaussian Hermitian random matrix ensemble, where $q = n =
1$) but has a different form if $q > n$ (which is of relevance to the
DoS {\it two-point} correlation function of a generic Gaussian
Hermitian random matrix ensemble, where $q = 2$, $n = 1$).
Establishing and interpreting this peculiar fact is the main message
of this paper.

The structure of the paper is as follows. In Section \ref{sect:2} we
derive the super-bosonization formula and the corresponding
supermatrix field theory. Our detailed treatment makes it clear that
the formula changes as one goes from one-point functions (i.e., the
mean density of states, DoS), to two-point functions (e.g., the
DoS-DoS correlation functions), to three-point functions, and so on.
In Section \ref{sect:almost} we use the formalism to compute
corrections to the DoS of the Gaussian ensemble of almost diagonal
random matrices. These were obtained earlier in Ref.\
\cite{YevKr2004} by another method. In Section \ref{sect:reduction}
we show how the new supermatrix field theory can be reduced to the
conventional NL$\sigma$M. In Section \ref{sect:conclusion} we give a
summary of the main results and outline some open problems. In
particular we discuss the possibility of a unified description of the
cases $q > n$ and $q\leq n\,$.

\section{The superbosonization formula}\label{sect:2}

The first few steps of the method presented here are similar to
previous derivations of the non-linear sigma model for random
matrices \cite{verbaarschot,efetov}. However, while the usual scheme
employs a Hubbard-Stratonovich transformation after the initial step
of averaging over the random matrix, here we will follow a different
route avoiding the Hubbard-Stratonovich transformation.

\subsection{Review of basic steps}

We begin with a generating function $Z_0\left[J\right]$,
\begin{equation}\label{eq:action1}
    Z_0[J] = \int D\bar\psi D\psi \, \exp \left(- \mathrm{i}
    \sum\nolimits_{i,j} \bar{\psi}_{i}\, s {\mathcal{H}}_{ij}^J
    \,\psi_{j}\right)\;, \qquad {\mathcal{H}}_{ij}^J = H_{ij} -
    \delta_{ij} \left( \mathcal{E} + J_{i}\right) \;,
\end{equation}
where $\psi_{i}$ are supervectors (with components arranged as column
vectors), $\bar\psi_{i}$ are the conjugate row supervectors, $H_{ij}$
are the matrix elements of our random matrix $H$, $J_{i}$ is the
source field, and $\mathcal{E}$ is a diagonal matrix containing the
energy parameters of the problem. The diagonal matrix $s$ with
entries $\pm 1$ keeps track of whether the Green's functions to be
computed from the generating function (\ref{eq:action1}) are retarded
(i.e., have an energy parameter with positive imaginary part) or
advanced (negative imaginary part). A scalar product or sum over
components $\bar\psi_i\, s\, \psi_j \equiv \sum_\lambda
\bar\psi_i^\lambda s^{\lambda} \psi_j^\lambda$ in the exponential is
understood. The number of components of the supervectors $\psi_i$ is
chosen according to the physical quantity under consideration. In the
most general situation each $\psi_i$ has $p$ anti-commuting and $q$
commuting components. To calculate such quantities as the two-level
correlation function one sets $p = q = 2\,$. In the case of the
density of states it suffices to use $p = q = 1\,$.

In order to make the presentation self-contained, we now give more
details concerning the supervectors $\psi_i$ and the matrix $s\,$.
The definitions are the same for every site, so we temporarily drop
the site index $i$ from our notation. We then have $\psi =
\begin{pmatrix} \chi \\ S \end{pmatrix}$ and $\bar\psi =
\begin{pmatrix} \bar\chi & \bar{S} \end{pmatrix}$ where
\begin{displaymath}
    \chi = \begin{pmatrix} \chi^1 \\ \vdots \\ \chi^p \end{pmatrix}
    \;, \quad S = \begin{pmatrix} S^1 \\ \vdots \\ S^q \end{pmatrix}
    \;, \quad \bar\chi = \begin{pmatrix} \chi^{1\ast} &\ldots
    &\chi^{p\ast} \end{pmatrix} \;, \quad \bar{S} = \begin{pmatrix}
    S^{1\ast} &\ldots &S^{q\ast} \end{pmatrix} \;,
\end{displaymath}
and $\chi^\mu$ and $S^\nu$ are anti-commuting and commuting
variables, respectively. $S^{\nu\ast}$ is the complex conjugate of
the complex variable $S^\nu$. The integral (\ref{eq:action1}) is
carried out w.r.t.\ the Berezin superintegral form
\begin{displaymath}
    D\bar\psi D\psi \equiv (2\pi)^{-p}\,d\bar{S} dS\,\partial_{\bar\chi}
    \partial_\chi \;, \quad d\bar{S} dS = \prod_{\nu=1}^q 2\,
    d \mathfrak{Re}(S^\nu) \, d\mathfrak{Im}(S^\nu) \;, \quad
    \partial_{\bar\chi} \partial_\chi = \prod_{\mu=1}^p \frac{
    \partial^2}{\partial\chi^{\mu\ast} \partial\chi^\mu} \;,
\end{displaymath}
with the domain of integration for each commuting variable $S^\nu$
being the complex numbers, $\mathbb{C}\,$. The integration measure
for the multi-site problem (\ref{eq:action1}) is of course $D\bar\psi
D\psi = \prod_i D\bar\psi_i D\psi_i$ (index $i$ reinstated). To
specify the sign matrix $s\,$, let us agree that we want $q_+$
retarded and $q_-$ advanced Green's functions (where $q_+ + q_- =
q)$. Then we may choose $s$ to be the diagonal matrix
\begin{displaymath}
    s = \mathrm{diag}( \mathbf{1}_p \, , \mathbf{1}_{q_+} \, ,
    - \mathbf{1}_{q_-} ) \;.
\end{displaymath}
(Note that there is absolutely no gain from introducing unnecessary
minus signs in the fermion-fermion sector.)

The next step is to take the average over the random matrix using the
Gaussian distribution
\begin{displaymath}
    \langle \ldots \rangle_{H} \equiv {\mathcal{N}} \int
    dH \, (\ldots)\, \exp \left( -{\textstyle\frac{1}{2}}\sum
    \nolimits_{i,j} \left\vert H_{ij} \right\vert^2 / C_{ij}\right)\;,
\end{displaymath}
where the coefficients $C_{ij}$ are real, non-negative, and symmetric
under exchanging $i \leftrightarrow j\,$, and ${\mathcal{N}}$ is the
normalization constant ensuring $\langle 1\rangle_H = 1\,$. Of course
$H_{ij}^\ast = H_{j i}$ by Hermiticity. The integration measure is
\begin{displaymath}
    dH := \prod_k dH_{kk} \prod_{i < j} 2\, d\mathfrak{Re}(H_{ij})
    \, d\mathfrak{Im}(H_{ij}) \;,
\end{displaymath}
where we are using a normalization convention which will be in force
throughout the paper.

An equivalent way of characterizing such a Gaussian distribution is
by means of its Fourier transform with respect to some arbitrary
commuting parameters $K_{ij}:$
\begin{equation}\label{e4}
    \left\langle \exp \left( - \mathrm{i} \sum\nolimits_{i,j} H_{ij}
    K_{j i} \right) \right\rangle_H = \exp \left( - {\textstyle{
    \frac{1}{2}}} \sum\nolimits_{i,j} C_{ij} K_{ij} K_{ji} \right) \;.
\end{equation}

The supervector integral representation (\ref{eq:action1}) of the
generating function $Z_{1}[J]$ allows us to average immediately over
the random matrix $H\,$. Using Eq.\ (\ref{e4}) for $K_{ji} \equiv
\bar\psi_i\, s\, \psi_j\,$, the average $Z[J]$ of the generating
function is
\begin{equation}\label{eq:Zpsi}
    Z[J] \equiv \left\langle Z_0[J] \right\rangle_H = \int D\bar\psi
    D\psi \, \exp \left( \mathrm{i}\sum\nolimits_i \bar\psi_i
    \, s\, (\mathcal{E} + J_i) \psi_i - {\textstyle\frac{1}{2}}
    \sum\nolimits_{i,j} C_{ij} (\bar\psi_i\, s\, \psi_j)
    (\bar\psi_j s\, \psi_i) \right)\;.
\end{equation}

Up to now, everything was more or less standard. In the next step,
however, we shall depart from the standard method \cite{efetov},
where one proceeds by making a Hubbard-Stratonovich transformation.
Instead, we will carry out a certain change of integration variables.
The change-of-variables formula to be introduced below bears some
resemblance to the well-known Weyl integration formula, by which one
reduces the integral over a conjugation-invariant function of a
matrix to an integral over the eigenvalues of the matrix.

{}From the cyclic property of the supertrace, one has the identities
\begin{displaymath}
    \mathrm{STr}\, (\psi_i \otimes \bar\psi_i) s =
    \bar\psi_i\, s\, \psi_i \;, \quad
    \mathrm{STr}\, (\psi_i \otimes \bar\psi_i) s\, (\psi_j
    \otimes \bar\psi_j) s = (\bar\psi_i\, s\, \psi_j)
    (\bar\psi_j s\, \psi_i)\;.
\end{displaymath}
Reading these identities backwards, one reorganizes the exponential
of the integrand in (\ref{eq:Zpsi}) as
\begin{displaymath}
    \mathrm{i}\sum\nolimits_i \mathrm{STr} \, (\psi_i \otimes
    \bar\psi_i) s\,( \mathcal{E} + J_i) - {\textstyle\frac{1}{2}}
    \sum\nolimits_{i,j} C_{ij}\, \mathrm{STr}\, (\psi_i \otimes
    \bar\psi_i) s\, (\psi_j \otimes \bar\psi_j) s \;,
\end{displaymath}
which prompts a change of integration variables
\begin{equation}\label{eq:var-change}
    \psi_i \otimes \bar\psi_i \to Q_i \;,
\end{equation}
where $Q_i$ is a supermatrix. Note that while the sign matrix $s$
does appear in the integrand, it will have \emph{no} influence on the
definition of the integration variable $Q_i\,$. Thus the domain of
integration for $Q_i$ will be independent of the types of Green's
functions (retarded or advanced) which are being generated. This is a
major difference from the traditional method using the
Hubbard-Stratonovich transformation.

\subsection{Heuristic approach}

The change of variables $\psi_i \otimes \bar\psi_i \to Q_i$ works in
the same way at every site $i\,$. Let us therefore focus on a fixed
site and simplify the notation by dropping the index $i$ for now.

In past work on the subject the desired change of variables
(\ref{eq:var-change}) was brought about by an unprecedented and
unexplained method, in the context of supermatrices, involving the
$\delta$-function. The logic went something like this: to transform a
Berezin superintegral such as $\int D\bar\psi D\psi \, f(\psi \otimes
\bar\psi)$, you insert
\begin{displaymath}
    1 \stackrel{?}{=} \int DQ \, \delta(Q - \psi \otimes \bar\psi)
\end{displaymath}
under the integral sign. Here $Q$ is a supermatrix with an
unspecified domain of integration. Then you reverse the order of
integration of $Q$ and $\psi$ to write
\begin{eqnarray*}
    &&\int D\bar\psi D\psi \, f(\psi \otimes \bar\psi) = \int DQ \,
    J(Q) \, f(Q) \;,\\ &&J(Q) \stackrel{?}{=} \int D\bar\psi D\psi \,
    \delta(Q - \psi\otimes\bar\psi) \stackrel{?}{=} \mathrm{SDet}(Q)\;.
\end{eqnarray*}

This kind of manipulation with the $\delta$-function requires
justification, which has never been provided. To give a simple
indication of what the issue is, consider the fermion-fermion sector
for the case of the two-level correlation function, i.e., put $p = 2$
and $q = 0\,$, in which case we are dealing with $2 \times 2$
matrices
\begin{displaymath}
    \chi \otimes \bar\chi =
    \begin{pmatrix} \chi^1 \chi^{1\ast} &\chi^1 \chi^{2 \ast} \\
    \chi^2 \chi^{1 \ast} &\chi^2 \chi^{2 \ast} \end{pmatrix} \;,
    \quad Q = \begin{pmatrix} Q^{11} &Q^{12} \\ Q^{21} &Q^{22}
    \end{pmatrix} \;.
\end{displaymath}
For concreteness let us say that $f(\psi \otimes \bar\psi) = f(\chi
\otimes \bar\chi) = \chi^1 \chi^{1 \ast} \chi^2 \chi^{2 \ast}$. Now,
in attempting to decide what function $f(Q)$ is to be placed in the
integrand we encounter an ambiguity. If we group the anti-commuting
variables as $f = (\chi^1 \chi^{1 \ast}) (\chi^2 \chi^{2 \ast})$ we
may be inclined to choose $f(Q) = Q^{11} Q^{22}\,$, but it is equally
valid to reorder the anti-commuting variables as $f = - (\chi^1
\chi^{2 \ast}) (\chi^2 \chi^{1 \ast})$, which would suggest $f(Q) = -
Q^{12} Q^{21}$. Perhaps a linear combination $f(Q) = Q^{11} Q^{22} -
t (Q^{11} Q^{22} + Q^{12} Q^{21})$ (with $t \in \mathbb{C}$) is the
good choice to make. All of these do the necessary job of returning
the given function $f = \chi^1 \chi^{1 \ast} \chi^2 \chi^{2 \ast}$ on
making the substitution $Q \to \chi \otimes \bar\chi\,$. Thus, unless
the matrix $Q$ is constrained by $Q^{11} Q^{22} + Q^{12} Q^{21}
\equiv 0\,$, an arbitrary parameter $t$ enters into the calculation.

At this stage it must be recalled that in mathematics there exists no
such thing as the $\delta$-\emph{function}. What does exist and can
be made sense of is the $\delta$-\emph{distribution}, say $\delta_x
\,$, which is the linear functional that evaluates test functions at
the point $x:$
\begin{displaymath}
    f \mapsto \delta_x[f] \equiv \int \delta(x-y)\, f(y)\, dy
    := f(x) \;.
\end{displaymath}
In our example, however, we are dealing with the element $f = \chi^1
\chi^{1 \ast} \chi^2 \chi^{2 \ast}$ of a Grassmann algebra -- not
with a function defined on points. It is unclear how we can evaluate
a $\delta$-distribution over these variables. In other words, the
meaning behind
\begin{displaymath}
    \delta(Q - \chi \otimes \bar\chi) \stackrel{?}{=}
    \delta(Q^{11} - \chi^1 \chi^{1 \ast})
    \, \delta(Q^{22} - \chi^2 \chi^{2 \ast})
    \, \delta(Q^{12} - \chi^1 \chi^{2 \ast})
    \, \delta(Q^{21} - \chi^2 \chi^{1 \ast}),
\end{displaymath}
viewed as a 'function' of the Grassmann variables, is undefined. No
meaning exists in current mathematics.

In the simple case under consideration, it is not difficult to work
out everything by hand and achieve control of the situation. Let us
take the integration domain for the $2 \times 2$ matrix $Q$ to be the
unitary group $\mathrm{U}(2)$ -- or any other 4-dimensional closed
submanifold of $\mathrm{GL}_2(\mathbb{C})$ in the same homology class
as $\mathrm{U}(2)$. Then if $d\mathrm{vol}_2(Q)$ denotes a Haar
measure on $\mathrm{U}(2)$, it can be shown that
\begin{displaymath}
    \int_{\mathrm{U}(2)} d\mathrm{vol}_2(Q)\, J(Q)\, ( Q^{11}Q^{22}
    + Q^{12} Q^{21}) = 0\;, \qquad J(Q) = \mathrm{Det}^{-1}(Q)\;.
\end{displaymath}
So, although bosonization of $f = \chi^1 \chi^{1 \ast} \chi^2 \chi^{2
\ast}$ is an ambiguous procedure leading to any one of the
one-parameter family of functions $F_t(Q) = Q^{11} Q^{22} - t (Q^{11}
Q^{22} + Q^{12} Q^{21})$, the ambiguity, at least in this case,
disappears at the level of integration provided we use $J(Q) =
\mathrm{Det}^{-1}(Q)$ and the integration domain $\mathrm{U}(2)$.

This does not solve the problem of what to do when the Grassmann
variables are more numerous or when considering the full
supersymmetric situation which is rendered more complicated by the
presence of commuting variables. It is clear that the same ambiguity
occurs in the general case: for any given function $f(\psi \otimes
\bar\psi)$, there exist many choices of function $F(Q)$ such that
$F(Q) \vert_{Q \to \psi \otimes \bar\psi}$ is equal to $f(\psi
\otimes \bar\psi)$. Given all these possible functions $F(Q)$ with
which the superbosonization formula
\begin{displaymath}
    \int D\bar\psi D\psi f(\psi \otimes \bar\psi) =
    \int DQ \, J(Q)\, F(Q) \;
\end{displaymath}
appears to be true, we may think we can use any one of them, or we
may attempt to impose some constraint on $Q$. Fortunately, recent
mathematical work \cite{ELSZ} has given a complete solution to this
problem for a restricted range of parameter values. In the next
subsection, we will present a summary of these results. Afterwards,
will give a detailed proof of the superbosonization formula for a
special but important case.

\subsection{Rigorous result}\label{sect:rigor}

In order to control the mathematics, let us assume that there is a
block or granular structure in the variances $C_{ij}\,$. By this we
mean if $i$ is a multi-index $i = (I,a)$ where $a = 1, \ldots, n\,$,
then
\begin{displaymath}
    C_{ij} =  C_{I,a\, ;I^\prime, a^\prime} = c_{I I^\prime}
    \delta_{a a^\prime}\;,
\end{displaymath}
independent of $a, a^\prime$. If so, then after grouping the terms
appropriately, the integrand depends only on the sums $\sum_{a=1}^n
\psi_{I,a} \otimes \bar\psi_{I^\prime,a}\,$. In this way the integer
$n$ (sometimes referred to as the number of orbitals) is introduced
as an additional parameter of our problem. Results stated in this
subsection are valid only in the range $n \ge q\,$. Later, we will
suggest how to recover the important case of $n=1$ when $q>1\,$.

To keep the notation simple, we return to using a single index $i$ as
a label for our supervectors $\psi_i\,$. In keeping with the
discussion above, we consider functions $f$ of the sum $\sum_{i=1}^n
\psi_i \otimes \bar\psi_i\,$. Our goal is to transform (by
superbosonization) the integral of such a function,
\begin{equation}\label{eq:inv-integral}
    \int f \equiv \int D\bar\psi D\psi\, f\left( \sum_{i=1}^n \psi_i
    \otimes \bar\psi_i \right) \;, \quad
    D\bar\psi D\psi \equiv \prod_{i=1}^n D\bar\psi_i D\psi_i \;.
\end{equation}

Let us address the general case of supervectors $\psi_i =
\begin{pmatrix} \chi_i \\ S_i \end{pmatrix}$ that have $p$
anti-commuting and $q$ commuting components (for $\chi_i$ and $S_i$
respectively). Then $\sum_{i=1}^n \psi_i \otimes \bar\psi_i$
corresponds to a supermatrix
\begin{displaymath}
    Q = \begin{pmatrix} A &\sigma\\ \tau &B \end{pmatrix} \;,
\end{displaymath}
where the blocks $A$ and $B$ are square matrices of size $p \times p$
and $q \times q$ with commuting variables as entries; while $\sigma$
and $\tau$ are rectangular matrices of size $p \times q$ and $q
\times p$ respectively and these have anti-commuting entries.

Let $F$ now be any function of the supermatrix $Q$ so that on making
the substitution $Q \to \sum \psi_i \otimes \bar\psi_i$ the function
$F(Q)$ becomes equal to the given function $f(\sum \psi_i \otimes
\bar\psi_i)$. We also assume that $n \ge q\,$. Then we claim that the
following equality of integrals holds:
\begin{equation}\label{bosonize}
    \int D\bar\psi D\psi\, f\left( \sum_{i=1}^n \psi_i \otimes
    \bar\psi_i \right) = \frac{\mathrm{vol}\, \mathrm{U}(n)}
    {\mathrm{vol}\, \mathrm{U}(n+p-q)} \int_D DQ \,
    \mathrm{SDet}^n(Q)\, F(Q)\;,
\end{equation}
provided $f$ decreases sufficiently fast at infinity so that the
integral on the left-hand side exists.

We now define all terms on the right-hand side of the identity
(\ref{bosonize}). The domain of integration $D$ is the unitary group
$\mathrm{U}(p)$ for $A$ and is the positive Hermitian $q \times q$
matrices, $\mathrm{Herm}^+(q)$, for $B\,$. Thus $D = \mathrm{U}(p)
\times \mathrm{Herm}^+(q)$. The Berezin superintegral form $DQ$ is
given by
\begin{equation}\label{eq:DEF-DQ}
    DQ :=  (2\pi)^{-pq}\, d\mathrm{vol}_p(A)
    \, d\mathrm{vol}_q(B) \, \partial_\tau \partial_\sigma \,
    \mathrm{Det}^q(A - \sigma B^{-1} \tau) \, \mathrm{Det}^p
    (B - \tau A^{-1} \sigma) \;,
\end{equation}
where $d\mathrm{vol}_p(A)$ is a Haar measure for $\mathrm{U}(p)$ and
$d\mathrm{vol}_q(B)$ is the correspondingly normalized invariant
measure for $\mathrm{Herm}^+(q)$. Invariance of $d\mathrm{vol}_q(B)$
means invariance under the transformation $B \mapsto g B g^\dagger$
for any $g \in \mathrm{GL}_q(\mathbb{C})$; the explicit expression
for $d\mathrm{vol}_q(B)$ is
\begin{displaymath}
    d\mathrm{vol}_q(B) = \frac{dB}{\mathrm{Det}^q(B)} \;, \quad
    dB \equiv \prod_{\lambda=1}^q dB^{\lambda\lambda}
    \prod_{1 \le \nu < \nu^\prime \le q} 2\, d\mathfrak{Re}
    (B^{\nu\nu^\prime})\, d\mathfrak{Im}(B^{\nu \nu^\prime})\;.
\end{displaymath}
The symbol $\partial_\tau \partial_\sigma$ is short for the product
of partial derivatives
\begin{displaymath}
    \partial_\tau \partial_\sigma = \prod_{\mu=1}^p \prod_{\nu=1}^q
    \frac{\partial^2}{\partial\tau^{\nu\mu}\,\partial\sigma^{\mu\nu}}\;.
\end{displaymath}
We mention in passing that the Berezin form (\ref{eq:DEF-DQ}) is
invariant under the action of a Lie superalgebra $\mathfrak{gl}(p|q)
\times \mathfrak{gl}(p|q)$ where the first factor acts on the left of
$Q$ and the second one on the right. This invariance property
actually determines $DQ$ uniquely up to multiplication by a constant.
The supermanifold of the supermatrix $Q$ belongs to one of the ten
families of Riemannian symmetric superspaces described in
\cite{suprev}. We also note that in the important case of $p = q$ the
Berezin form (\ref{eq:DEF-DQ}) is flat, i.e., is given by a product
of differentials resp.\ partial derivatives:
\begin{equation}\label{eq:def-dq}
    DQ = (2\pi)^{-p^2} |dA| \, dB \, \partial_\tau \partial_\sigma\;.
\end{equation}
The superdeterminant is the usual one, satisfying $\ln \mathrm{SDet}
(Q) = \mathrm{STr}\, \ln Q$ with $\mathrm{STr}\, Q = \mathrm{Tr}\, B
- \mathrm{Tr}\, A$ (bosons count as plus, fermions as minus):
\begin{equation}\label{eq:def-SDet}
    \mathrm{SDet}^n(Q) = \frac{\mathrm{Det}^n(B)}{\mathrm{Det}^n
    (A - \sigma B^{-1} \tau )} \;.
\end{equation}
We repeat once more that the inequality $n \ge q$ has to be satisfied
in order for the formula (\ref{bosonize}) to be true.

For the case that $f$ is a Schwartz function (i.e., decreases faster
than any power), a mathematical proof of the superbosonization
formula (\ref{bosonize}) has been given in \cite{ELSZ}. More
precisely, the theorem stated and proved in \cite{ELSZ} assumes
(besides $n \ge q$) that $f$ is a \emph{holomorphic} $\mathrm{U}
(n)$-invariant function of the vectors $S_i\,$, $\bar{S}_i\,$,
$\chi_i\,$, $\bar\chi_i$ $(i = 1, \ldots, n)$. These assumptions are
always satisfied for the case of Gaussian disorder distributions.

In summary, compared to previous works which suggest the existence a
formula such as (\ref{bosonize}), the main advance here is that the
integration domain and sufficient conditions of validity have been
specified. In the light of the discussion in the previous subsection,
let us emphasize once more that there exist many choices of function
$F$ (although in practical applications there will usually be a
natural choice) so that $F(Q)$ becomes $f(\sum \psi_i \otimes
\bar\psi_i)$ on substituting $Q \to \psi_i \otimes \bar\psi_i\,$. The
theorem in \cite{ELSZ} states that the superbosonization formula
(\ref{bosonize}) holds true for \emph{any} such choice of $F$ and
that paper also gives analogous formulas for the cases of orthogonal
and symplectic symmetry.

\subsection{Proof of the superbosonization formula for $p = q = n$}
\label{sect:p=q=n}

We now elaborate on the case of $p = q = n$, which will be a good
starting point for making the generalization to $n < q$ in Section
\ref{sect:KY}. An application of the special case of $p = q = n = 1$
will be given in Section \ref{sect:almost}.

We simplify the notation by putting $r := p = q = n$. Thus we are now
dealing with $r$ vectors $S_1 , \ldots, S_r$ each of which has $r$
complex components, and with $r$ vectors $\chi_1 , \ldots, \chi_r$
having $r$ anti-commuting components. Let $f$ be some function that
depends only on the $\mathrm{U}(n)$-invariant combinations $(n = r)$
\begin{displaymath}
    \sum_{i=1}^r \chi_i \otimes \bar\chi_i \;, \quad
    \sum_{i=1}^r \chi_i \otimes \bar{S}_i \;, \quad
    \sum_{i=1}^r S_i \otimes \bar\chi_i \;, \quad
    \sum_{i=1}^r S_i \otimes \bar{S}_i \;,
\end{displaymath}
viewed as the blocks of a supermatrix. We wish to compute the
integral of such a function:
\begin{equation}\label{eq:18}
    \int f \equiv (2\pi)^{-r^2} \int d\bar{S} dS \, \partial_{\bar\chi}
    \partial_\chi \, f \begin{pmatrix} \sum_i \chi_i \otimes \bar\chi_i
    & \sum_i \chi_i \otimes \bar{S}_i \\ \sum_i S_i \otimes \bar\chi_i
    &\sum_i S_i \otimes \bar{S}_i \end{pmatrix} \;.
\end{equation}
For this purpose we choose some function $F(Q)$ of a supermatrix $Q$
so that making the substitution
\begin{displaymath}
    Q \to \sum_i \psi_i \otimes \bar\psi_i =
    \begin{pmatrix} \sum_i \chi_i \otimes \bar\chi_i & \sum_i \chi_i
    \otimes \bar{S}_i \\ \sum_i S_i \otimes \bar\chi_i &\sum_i S_i
    \otimes \bar{S}_i \end{pmatrix} \;,
\end{displaymath}
we recover from $F$ the given function $f$. In the following we will
change the integral $\int f$ in (\ref{eq:18}) to an integral over $F$
in four steps.

\subsubsection{First step}\label{sect:step-one}

We begin by assuming that the vectors $S_1 , \ldots, S_r$ are
linearly independent, so that the $r \times r$ matrix which is formed
by taking the vectors $S_1, \ldots, S_r$ to be the columns of that
matrix is regular or of full rank. (The condition of linear
independence is not always satisfied, of course. However, the sets of
linearly dependent vectors form a set of measure zero. They can
therefore be ignored for the purpose of integration.) Then we define
square matrices $\sigma$ and $\tau$ of size $r \times r$ and with
anti-commuting entries by
\begin{displaymath}
    \sigma := \sum_{i = 1}^r \chi_i \otimes \bar{S}_i \;, \quad
    \tau := \sum_{i = 1}^r S_i \otimes \bar\chi_i \;.
\end{displaymath}
By the chain rule of differentiation we have
\begin{displaymath}
    \partial_{\bar\chi} \partial_\chi = \mathrm{Det}^r \left( \sum S_i
    \otimes \bar{S}_i \right)\, \partial_{\tau} \partial_{\sigma}\;,
\end{displaymath}
where $\partial_\tau \partial_\sigma \equiv \prod \partial^2 /
\partial \tau^{\nu\mu} \partial\sigma^{\mu\nu}\,$. By the assumption
that the vectors $S_1 , \ldots, S_r$ are linearly independent, the $r
\times r$ matrix
\begin{displaymath}
    B := \sum_{i=1}^r S_i \otimes \bar{S}_i
\end{displaymath}
has non-vanishing determinant and hence an inverse. Thus we can write
\begin{displaymath}
    \sum_{i=1}^r \chi_i \otimes \bar\chi_i = \left( \sum \chi_i \otimes
    \bar{S}_i \right) \left( \sum S_j \otimes \bar{S}_j \right)^{-1}
    \left(\sum S_k \otimes \bar\chi_k \right) = \sigma B^{-1} \tau \;,
\end{displaymath}
and our integral (\ref{eq:18}) then becomes
\begin{displaymath}
    \int f = (2\pi)^{-r^2} \int d\bar{S} dS \, \mathrm{Det}^r(B) \,
    \partial_\tau \partial_\sigma \, F \begin{pmatrix} \sigma B^{-1}
    \tau & \sigma \\ \tau &B \end{pmatrix} \;.
\end{displaymath}

\subsubsection{Second step}\label{sect:step-two}

Now we take the Hermitian (and non-negative) matrix $B = \sum S_i
\otimes \bar{S}_i$ as the new set of commuting variables of
integration. Let $\delta(B)$ be the $\delta$-distribution (or
$\delta$-function) centered at zero on the linear space of Hermitian
$r \times r$ matrices $B$, and let $\delta(B - \sum S_i \otimes
\bar{S}_i)$ be the shifted $\delta$-distribution. We then have
\begin{displaymath}
    1 = \int dB \, \delta \left( B - \sum S_i \otimes
    \bar{S}_i \right) \;,
\end{displaymath}
where the integral is over the linear space of Hermitian $r \times r$
matrices $B$ with flat (or translation-invariant) measure $dB =
\prod_k dB_{kk} \prod_{i < j} 2\, d\mathfrak{Re}(B_{ij}) \, d
\mathfrak{Im}(B_{ij})$. Now if $f_1$ is an (integrable) function of
$\sum S_i \otimes \bar{S}_i\,$, then by inserting the relation above
and changing the order of integration we get
\begin{eqnarray*}
    \int d\bar{S} dS \, f_1 \left( \sum S_i \otimes \bar{S}_i \right)
    &=& \int d\bar{S} dS \, \left( \int dB \, \delta \left( B -
    \sum S_i \otimes \bar{S}_i \right) \right) \, f_1 \left( \sum S_i
    \otimes \bar{S}_i \right) \\ &=& \int dB \, f_1(B) \int d\bar{S}
    dS\, \delta \left( B - \sum S_i \otimes \bar{S}_i \right) \;.
\end{eqnarray*}
To compute the integral $J(B) := \int d\bar{S} dS\, \delta \left( B -
\sum S_i \otimes \bar{S}_i \right)$, let $g \in \mathrm{GL}_r
(\mathbb{C})$ be any invertible $r \times r$ matrix with complex
entries, and consider
\begin{displaymath}
    J(g B g^\dagger) = \int d\bar{S} dS\, \delta \left(
    g B g^\dagger - \sum S_i \otimes \bar{S}_i \right) \;.
\end{displaymath}
Of course, since $\sum S_i \otimes \bar{S}_i \ge 0\,$, the function
$J(B)$ vanishes if one or several eigenvalues of $B$ are negative.
Changing variables from $S_i$ to $g S_i$ ($i = 1, \ldots, r$) gives
\begin{displaymath}
    J(g B g^\dagger) = \int d\bar{S} dS \, \mathrm{Det}^r (g g^\dagger)
    \, \delta \left( g \left( B - \sum S_i \otimes \bar{S}_i \right)
    g^\dagger \right) \;.
\end{displaymath}
Now the $\delta$-function obeys the transformation rule $\delta(g B
g^\dagger) = \mathrm{Det}^{-r}(g g^\dagger) \, \delta(B)$. We
therefore obtain $J(g B g^\dagger) = J(B)$. Since the action of
$\mathrm{GL}_r (\mathbb{C})$ on the positive Hermitian matrices by $B
\mapsto g B g^\dagger$ is transitive (i.e., every positive Hermitian
matrix $B$ can be written in the form $B = g g^\dagger$), it follows
that $J(B)$ is a constant independent of $B > 0\,$.

This constant is readily found by taking $B$ to be the $r \times r$
unit matrix, $B = \mathbf{1}_r\,$. Contributions to the integral
\begin{displaymath}
    J(\mathbf{1}_r) = \int d\bar{S} dS\, \delta \left(
    \mathbf{1}_r - \sum S_i \otimes \bar{S}_i \right)
\end{displaymath}
come only from those $S_1 , \ldots, S_r$ that form an orthonormal
system. Therefore $J(\mathbf{1}_r) = \mathrm{vol}\, \mathrm{U}(r)$.
More precisely, if we make the factorization $(S_1, S_2, \ldots, S_r)
= \mathrm{e}^X \mathrm{e}^{\mathrm{i} Y}$ with Hermitian $X$ and $Y$,
then $d\bar{S} dS$ becomes $dB \, dk$ where $B = \mathrm{e}^{2X}$ and
$dk$ for $k = \mathrm{e}^{\mathrm{i}Y}$ is the Haar measure for
$\mathrm{U}(r)$ normalized by
\begin{displaymath}
    dk \vert_{Y = 0} = dY \equiv \prod_l dY_{ll} \prod_{i
    < j} 2\, d\mathfrak{Re} (Y_{ij}) \, d\mathfrak{Im}(Y_{ij}) \;.
\end{displaymath}
Thus $J(\mathbf{1}_r) = \int dk \int dB \, \delta( \mathbf{1}_r - B)
= \mathrm{vol}\, \mathrm{U}(r)$. After some of calculations (passing
through the eigenvalue representation of $k$) one can show that with
this choice of normalization the total volume of $\mathrm{U}(r)$ is
\begin{displaymath}
    \mathrm{vol}\, \mathrm{U}(r) = \int\limits_{\mathrm{U}(r)} dk
    = \frac{(2\pi)^{r(r+1)/2}}{0!\, 1!\, \cdots (r-1)!} \;.
\end{displaymath}
Altogether we obtain
\begin{displaymath}
    \int d\bar{S} dS \, f_1\left( \sum S_i \otimes \bar{S}_i \right) =
    \mathrm{vol}\, \mathrm{U}(r) \int\limits_{B>0} dB \, f_1(B) \;.
\end{displaymath}
Applying this relation to our situation gives
\begin{displaymath}
    \int f = (2\pi)^{-r^2} \mathrm{vol}\, \mathrm{U}(r)
    \int\limits_{B > 0} dB \, \mathrm{Det}^r(B)\,\partial_\tau
    \partial_\sigma F \begin{pmatrix} \sigma B^{-1} \tau &\sigma\\
    \tau &B \end{pmatrix} \;.
\end{displaymath}

\subsubsection{Third step}

We now re-express the volume factor $\mathrm{vol}\, \mathrm{U}(r)$ as
a dummy integral $\mathrm{vol}\, \mathrm{U}(r) = \int_{\mathrm{U}(r)}
dk\,$. Having done so, we view $\mathrm{U}(r)$ as a real subgroup of
the complex group $\mathrm{GL}_r(\mathbb{C})$, and we extend the Haar
measure $dk$ (regarded as a differential form) from $\mathrm{U}(r)$
to $\mathrm{GL}_r (\mathbb{C})$. For example, in the case of $r = 1$
this means that we set $k = \mathrm{e}^{\mathrm{i}y}$ with $y \in
[0,2\pi]$ and extend the given Haar measure $d y$ of the unit circle
$\mathrm{U}(1) \subset \mathbb{C}$ to the holomorphic differential
form $dy = (\mathrm{i} a)^{-1} da$ on $\mathrm{GL}_1(\mathbb{C})$ by
letting $a = \mathrm{e}^{\mathrm{i}y}$ be any non-zero complex
number.

In the general case, if $A_{ij}$ are the matrix elements of $A \in
\mathrm{GL}_r(\mathbb{C})$, the Haar measure $dk$ extends from
$\mathrm{U}(r)$ to $\mathrm{GL}_r(\mathbb{C})$ as the holomorphic
differential form
\begin{displaymath}
    dk = \mathrm{i}^{-r^2} \frac{dA}{\mathrm{Det}^r(A)} \;,
    \quad dA = \bigwedge_{i,j} dA_{i,j} \;,
\end{displaymath}
where the variables $A_{ij}$ for $i , j = 1, \ldots, r$ are regarded
as a set of $r^2$ independent complex coordinates subject only to the
condition $\mathrm{Det}(A) \not= 0$. (Mathematically speaking, by
interpreting the Haar measure as a differential form we choose a
fixed orientation of the unitary group.) To verify this formula for
$dk$, notice that $\mathrm{Det}^{-r}(A)\, dA$ is invariant under the
transformation $A \mapsto g A h$ for any $g , h \in \mathrm{GL}_r(
\mathbb{C})$. This invariance property in fact determines the Haar
measure $dk$ uniquely up to multiplication by a constant. To verify
the normalization constant, one sets $A = \mathrm{e}^{ \mathrm{i} Y}$
and notices that $\mathrm{Det}^{-r}(A)\, dA \vert_{Y = 0} =
\mathrm{i}^{r^2} dY$.

Let us now assume that $F \begin{pmatrix} \sigma B^{-1} \tau &\sigma
\\ \tau &B \end{pmatrix}$ has been chosen as an analytic function of
the left upper block. Then, using the fact that
\begin{displaymath}
    \int_{\mathrm{U}(r)} dk \, k_{i_1,j_1} k_{i_2,j_2} \cdots
    k_{i_l,j_l} = 0
\end{displaymath}
holds for any $l > 0$, we can write
\begin{eqnarray*}
    \int f &=& (2\pi)^{-r^2} \int\limits_{B > 0} dB \, \mathrm{Det}^r
    (B)\int\limits_{\mathrm{U}(r)} dk \, \partial_\tau \partial_\sigma
    F \begin{pmatrix} k + \sigma B^{-1} \tau &\sigma \\
    \tau &B \end{pmatrix} \\ &=& (2\pi\mathrm{i})^{-r^2}
    \int\limits_{B > 0} dB \, \mathrm{Det}^r(B)
    \int\limits_{\mathrm{U}(r)} dA \, \mathrm{Det}^{-r}(A) \,
    \partial_\tau \partial_\sigma F \begin{pmatrix} A + \sigma B^{-1}
    \tau &\sigma \\ \tau &B \end{pmatrix} \;.
\end{eqnarray*}

\subsubsection{Fourth step}

In the final step, we use the fact that the holomorphic form $dA$ is
invariant under translations. We exploit this invariance to make the
(nilpotent) translation $A \to A - \sigma B^{-1} \tau$. Since the
unitary group is a closed manifold (i.e., has no boundary), such a
translation does not give rise to boundary terms. Therefore, the
result after translation is
\begin{displaymath}
    \int f = (2\pi\mathrm{i})^{-r^2} \int\limits_{B > 0} dB
    \int\limits_{\mathrm{U}(r)} dA \, \partial_\tau \partial_\sigma
    \frac{\mathrm{Det}^r(B)}{\mathrm{Det}^r(A - \sigma B^{-1} \tau)}
    \, F \begin{pmatrix} A &\sigma\\ \tau &B \end{pmatrix} \;.
\end{displaymath}
To bring this result into standard form, we write
\begin{displaymath}
    DQ := (2\pi)^{-r^2} \left( \frac{dB}{\mathrm{Det}^r(B)} \right)
    \left( \frac{dA}{\mathrm{i}^{r^2} \mathrm{Det}^r(A)} \right)
    \partial_\tau \partial_\sigma \mathrm{Det}^r(A)\,
    \mathrm{Det}^r(B) \;,
\end{displaymath}
where the second factor (returning from differential forms to
measures) agrees with our Haar measure $dk$ for $\mathrm{U}(r)$, and
the first factor is a correspondingly normalized invariant measure
for $\mathrm{Herm}^+(r)$. These were denoted in the more general
situation of Section \ref{sect:rigor} by $d\mathrm{vol}_p(A)$ and
$d\mathrm{vol}_q(B)$ respectively. Note that from the relation
\begin{displaymath}
    \mathrm{Det}^q(A-\sigma B^{-1} \tau)\, \mathrm{Det}^p(B - \tau
    A^{-1} \sigma) = \mathrm{Det}^r(A)\, \mathrm{Det}^r(B)
\end{displaymath}
for $p = q = r$, the present expression for $DQ$ in fact coincides
with (\ref{eq:DEF-DQ}).

Since the ratio of determinants $\mathrm{Det}(B) / \mathrm{Det}(A -
\sigma B^{-1} \tau)$ equals the superdeterminant of the supermatrix
$Q = \begin{pmatrix} A &\sigma \\ \tau &B \end{pmatrix}$, we now
arrive at the desired formula
\begin{displaymath}
    \int f = \int DQ \, \mathrm{SDet}^r(Q) F(Q) \;,
\end{displaymath}
where it is understood that we integrate over the unitary matrices $A
\in \mathrm{U}(r)$ and the positive Hermitian $r \times r$ matrices
$B \in \mathrm{Herm}^+(r)$. This completes the proof of the
superbosonization formula (\ref{bosonize}) for $p = q = n = r\,$.

\subsection{The case of $n < q$}\label{sect:KY}

We now turn to the case of $p$ fermionic replicas and $q > n$ bosonic
ones. This case is not covered by the results of \cite{ELSZ} and
needs separate treatment. As before, we wish to calculate the
integral (\ref{eq:inv-integral}).

Let us first verify by inspection that formula (\ref{bosonize})
cannot be true in the present situation. For this purpose, let $p = q
= r$ (but $r > n$) for simplicity, and consider some function $f$
that depends only on the combinations $\sum_{i=1}^n \chi_i \otimes
\bar\chi_i$ and $\sum_{i=1}^n S_i \otimes \bar{S}_i \,$. After
superbosonization the integrand $F$ is only a function of $A$ and $B$
(and does not depend on $\sigma$ and $\tau$). As we noted above, the
invariant Berezin form (\ref{eq:DEF-DQ}) for $p = q = r$ is
\begin{displaymath}
    DQ = (2\pi)^{-r^2} d\mathrm{vol}_r(A) \, d\mathrm{vol}_r(B)
    \, \partial_\tau \partial_\sigma \, \mathrm{Det}^r(A) \,
    \mathrm{Det}^r(B) \;.
\end{displaymath}
Also, using the expression (\ref{eq:def-SDet}) for the
superdeterminant, our integrand becomes
\begin{displaymath}
    DQ\,\mathrm{SDet}^n(Q)\,F(Q) = (2\pi)^{-r^2} d\mathrm{vol}_r(A)\,
    \mathrm{Det}^{r-n}(A)\, d\mathrm{vol}_r(B)\, \mathrm{Det}^{r+n}(B)
    \, \partial_\tau \partial_\sigma\, \mathrm{Det}^n
    \left(\mathbf{1}_r - B^{-1} \tau A^{-1} \sigma \right) \;.
\end{displaymath}
Now the Fermi integral
\begin{displaymath}
    \partial_\tau \partial_\sigma\, \mathrm{Det}^n
    \left(\mathbf{1}_r - B^{-1} \tau A^{-1} \sigma \right)
\end{displaymath}
vanishes identically, since $\mathrm{Det}^n (\mathbf{1}_r - B^{-1}
\tau A^{-1} \sigma)$ can be at most of degree $n \times 2r$ in the
matrix elements of $\sigma$ and $\tau$, whereas the differential
operator $\partial_\tau \partial_\sigma$ is homogeneous of higher
degree $2 r^2 > 2nr\,$. (Thus there are not enough anti-commuting
variables in the integrand to give a non-zero result when taking all
partial derivatives.)

However, if we scale $A$ and $B$ out of $\mathrm{Det}^n (\mathbf{1}_r
- B^{-1} \tau A^{-1} \sigma)$ by sending, say, $\sigma \to A \sigma$
and $\tau \to B \tau$, then the $B$-dependence of the integrand
becomes
\begin{displaymath}
    d\mathrm{vol}_r(B)\, \mathrm{Det}^n(B)\, F(\cdot,B)
    = dB \, \mathrm{Det}^{n-r}(B)\, F(\cdot,B) \;,
\end{displaymath}
which gives rise to a singularity when one (or several) of the
eigenvalues of the positive matrix $B$ approach zero. This
singularity is non-integrable if the integrand $F$ goes to a non-zero
constant in the same limit. Thus the right-hand side of the
superbosonization formula (\ref{bosonize}) is ill-defined (of the
type of $0 \times \infty$) in the present case. Based on the
treatment given for $n = r$ in Section \ref{sect:p=q=n}, we will now
derive the correct formula for $n < r\,$.

\subsubsection{First step}

We now have $n$ vectors $S_1, \ldots, S_n$, each with $q$ complex
components. We arrange these as a $q \times n$ rectangular matrix $S
:= (S_1 , \ldots, S_n)$ so that
\begin{displaymath}
    S S^\dagger = \sum_{i=1}^n S_i \otimes \bar{S}_i \;.
\end{displaymath}
We decompose the rectangular matrix $S$ as
\begin{displaymath}
    S = \begin{pmatrix} a \\ b \end{pmatrix} \;,
\end{displaymath}
where $a$ is an $n \times n$ square matrix while the block $b$ is
rectangular of size $(q-n) \times n\,$. For generic $S$ the square
matrix $a$ is invertible and hence is an element $a \equiv g$ of the
group $\mathrm{GL}_n(\mathbb{C})$. Defining $Z := b\, a^{-1}$ we then
have
\begin{displaymath}
    S = \begin{pmatrix} g \\ Z g \end{pmatrix} \;.
\end{displaymath}
By simple power counting one sees that the volume element transforms
as
\begin{displaymath}
    d\bar{S} dS = d\bar{Z} dZ \, d\bar{g} dg \, \mathrm{Det}^{q-n}(g
    g^\dagger) \;.
\end{displaymath}
Here our notational conventions are the same as before, i.e., $d
\bar{g} dg := \prod_{i,j=1}^n 2\, d\mathfrak{Re}(g_{ij})\, d
\mathfrak {Im} (g_{ij})$ and the same expression goes for $d\bar{Z}
dZ$. We now make a further change of variables $g \to (1 + Z^\dagger
Z)^{-1/2} h\,$, which results in
\begin{displaymath}
    S = \begin{pmatrix} (1 + Z^\dagger Z)^{-1/2} h \\
    Z (1 + Z^\dagger Z)^{-1/2} h \\ \end{pmatrix} \;, \quad
    d\bar{S} dS = \frac{d\bar{Z} dZ}{\mathrm{Det}^q (1 + Z^\dagger Z)}
    \, d\bar{h} dh \, \mathrm{Det}^{q-n} (h h^\dagger) \;.
\end{displaymath}

To explain the above factorization of $S$ let $\mathrm{Mat}^\prime
_{q,n}(\mathbb{C})$ denote all complex $q \times n$ matrices with
full rank $n\,$. Every $S \in \mathrm{Mat}^\prime_{q,n}(\mathbb{C})$
can be decomposed as $S = T h$ with $h \in \mathrm{GL}_n(\mathbb{C})$
and $T \in \mathrm{Mat}^ \prime_{q,n}(\mathbb{C})$ being the
truncation of a unitary $q \times q$ matrix to the first $n$ columns
(i.e., the last $q-n$ columns are deleted). This decomposition is not
unique. Indeed, if $S = T h$ is such a decomposition, then $S = (T
k^{-1}) (k h)$ is also such a decomposition for any $k \in
\mathrm{U}(n)$. Thus the correct mathematical statement of
factorization is
\begin{displaymath}
    \mathrm{Mat}^\prime_{q,n}(\mathbb{C}) = (\mathrm{U}(q) /
    \mathrm{U}(q-n))\times_{\mathrm{U}(n)} \mathrm{GL}_n(\mathbb{C})\;.
\end{displaymath}
Locally -- more precisely speaking: whenever $\mathrm{Det}^{-q}(1 +
Z^\dagger Z) \not= 0$ -- we can make a definite choice of $k$ by
taking
\begin{displaymath}
    T := \begin{pmatrix} (1 + Z^\dagger Z)^{-1/2} \\
    Z (1 + Z^\dagger Z)^{-1/2} \end{pmatrix} \in \mathrm{Mat}^\prime
    _{q,n}(\mathbb{C}) \;.
\end{displaymath}
The factorization $S = T h$ then means that we regard $S$ as being
given by its regular part $h \in \mathrm{GL}_n(\mathbb{C})$ and a
point of the Grassmann manifold $\mathrm{U}(q) / \mathrm{U}(q-n)
\times \mathrm{U}(n)$, which is the set of realizations of
$\mathbb{C}^n$ as a unitary subspace of $\mathbb{C}^q$.

{}From the standard fact
%
%
that the metric tensor of $\mathrm{U}(q) / \mathrm{U}(n) \times
\mathrm{U}(q-n)$ can be expressed by
\begin{displaymath}
    \mathrm{Tr}\, dZ (\mathbf{1}_n + Z^\dagger Z)^{-1}
    d Z^\dagger (\mathbf{1}_{q-n} + Z Z^\dagger)^{-1} \;,
\end{displaymath}
one easily finds that $d\bar{Z} dZ \, \mathrm{Det}^{-q}(1 + Z^\dagger
Z)$ expresses the invariant measure of $\mathrm{U}(q) / \mathrm{U}
(q-n) \times \mathrm{U}(n)$. We henceforth denotes this invariant
measure by $d\mathrm{vol}(T)$. Thus we can summarize the discussion
of this subsection by
\begin{displaymath}
    S = T h \;, \quad d\bar{S} dS = d\mathrm{vol}(T) \,
    d\bar{h} dh \, \mathrm{Det}^{q-n} (h h^\dagger) \;,
\end{displaymath}
which results in the formula
\begin{displaymath}
    \int f = (2\pi)^{-pn} \int d\mathrm{vol}(T) \int d\bar{h} dh\,
    \mathrm{Det}^{q-n}(h h^\dagger)\, \partial_{\bar\chi}\partial_\chi
    \, F \begin{pmatrix} \chi \bar\chi &\chi h^\dagger T^\dagger
    \\ T h \bar\chi &T h h^\dagger T^\dagger \end{pmatrix} \;.
\end{displaymath}
Here $\chi$ is the $p \times n$ rectangular matrix $\chi = (\chi_1 ,
\ldots, \chi_n)$, and $\bar\chi$ is the corresponding $n \times p$
rectangular matrix whose rows are the row vectors $\bar\chi_1 ,
\ldots, \bar\chi_n\,$.  In other words, we have $\sum \chi_i \otimes
\bar\chi_i = \chi \bar\chi\,$, $\sum \chi_i \otimes \bar{S}_i = \chi
S^\dagger$, etc.

\subsubsection{Second step}

In the next step we take the matrix elements of $\sigma := \chi
h^\dagger$ and $\tau := h \bar\chi$ as our new anti-commuting
variables. This gives (cf.\ Section \ref{sect:step-one})
\begin{displaymath}
    \int f = (2\pi)^{-pn} \int d\mathrm{vol}(T) \int d\bar{h} dh\,
    \mathrm{Det}^{p+q-n}(h h^\dagger)\, \partial_\tau \partial_\sigma\,
    F \begin{pmatrix} \sigma (h h^\dagger)^{-1} \tau &\sigma
    T^\dagger\\ T \tau &T h h^\dagger T^\dagger \end{pmatrix} \;.
\end{displaymath}
Then we make a change of (commuting) variables to $B := h h^\dagger$.
By the same reasoning given in detail in Section \ref{sect:step-two},
we obtain
\begin{displaymath}
    \int f = (2\pi)^{-pn}\, \mathrm{vol}\, \mathrm{U}(n) \int
    d\mathrm{vol}(T) \int\limits_{B > 0} dB\, \mathrm{Det}^{p+q-n}(B)
    \, \partial_\tau \partial_\sigma\, F \begin{pmatrix}\sigma B^{-1}
    \tau &\sigma T^\dagger\\ T \tau &T B T^\dagger \end{pmatrix}\;.
\end{displaymath}
The following steps are also similar to before. We introduce a dummy
integral over $\mathrm{U}(p)$ with $(\mathrm{vol}\, \mathrm{U}(p)
)^{-1} \times \int_{\mathrm{U}(p)} \mathrm{Det}^{-p}(\mathrm{i}A)\,dA
= 1:$
\begin{displaymath}
    \int f = (2\pi)^{-pn}\, \frac{\mathrm{vol}\, \mathrm{U}(n)}
    {\mathrm{vol}\, \mathrm{U}(p)} \int d\mathrm{vol}(T)
    \int\limits_{B > 0} dB \int\limits_{\mathrm{U}(p)} dA
    \, \frac{\mathrm{Det}^{p+q-n}(B)}{\mathrm{Det}^p(\mathrm{i}A)}
    \, \partial_\tau \partial_\sigma\, F \begin{pmatrix}
    A + \sigma B^{-1} \tau &\sigma T^\dagger\\ T \tau &T B T^\dagger
    \end{pmatrix} \;.
\end{displaymath}
Then we make the shift $A \to A - \sigma B^{-1} \tau$. Expressing the
result in terms of the invariant measures $d\mathrm{vol}_p(A) =
\mathrm{Det}^{-p}(\mathrm{i}A)\, dA$ and $d\mathrm{vol}_n(B) =
\mathrm{Det}^{-n}(B)\, dB\,$, we obtain
\begin{equation}\label{eq:KY-formula}
    \int f = (2\pi)^{-pn}\, \frac{\mathrm{vol}\, \mathrm{U}(n)}
    {\mathrm{vol}\, \mathrm{U}(p)} \int d\mathrm{vol}(T)
    \int\limits_{B > 0} d\mathrm{vol}_n(B)
    \int\limits_{\mathrm{U}(p)} d\mathrm{vol}_p(A)
    \, \partial_\tau \partial_\sigma\,
    \frac{\mathrm{Det}^q(B)}{\mathrm{Det}^{-p}(B -
    \tau A^{-1} \sigma)} F \begin{pmatrix}
    A &\sigma T^\dagger\\ T \tau &T B T^\dagger\end{pmatrix} \;.
\end{equation}
This formula is arguably more complicated than (\ref{bosonize}). The
reason for this is that the rank of $\sum_{i=1}^n S_i \otimes
\bar{S}_i$ never exceeds $n\,$, so there exist at least $q-n$ zero
eigenvalues and the range of $\sum S_i \otimes \bar{S}_i$ is a
\emph{submanifold of the boundary} of $\mathrm{Herm}^+(q)$. Our
derivation parameterizes this submanifold by two factors. The first
factor, $T$, describes the $n$-dimensional complement of the kernel
space of $\sum S_i \otimes \bar{S}_i$ in $\mathbb{C}^q$; thus it
describes the complex $n$-plane spanned by the vectors $S_1, \ldots,
S_n$ in $\mathbb{C}^q$. (The set of such subspaces $\mathbb{C}^n
\hookrightarrow \mathbb{C}^q$ is in one-to-one correspondence with
the symmetric space $\mathrm{U}(q) / \mathrm{U}(n) \times
\mathrm{U}(q-n)$). The second factor, $B$, is the operator $\sum S_i
\otimes \bar{S}_i$ restricted to its complex $n$-plane
$\mathbb{C}^n$.

The merit of the result (\ref{eq:KY-formula}) is it that allows us to
make an exact transformation of the original problem to supermatrix
variables in the parameter range $n < q\,$. Unfortunately, the
invariance properties of the integral on the right-hand side are not
very transparent. In other words, the Lie superalgebra
$\mathfrak{gl}(p|q) \times \mathfrak{gl}(p|q)$ acts as first-order
differential operators on the functions $f$ and $F$, and while the
transformation behavior with respect to this action is very clear on
the left-hand side, it is not easy to see how the desired behavior
emerges on the right-hand side.

\subsection{Supermatrix model}

Let us now return to the formulation of Section \ref{sect:rigor},
which is valid for $n \ge q\,$. The superbosonization formula
(\ref{bosonize}) allows us to replace the initial problem of
computing the correlation functions for a Gaussian ensemble of
Hermitian random matrices with variance matrix $C_{ij}\,$, Eq.\
(\ref{e4}), with the problem of computing the generating function
\begin{equation}
\begin{split}
    Z[J] &= \int \prod_{i=1}^M DQ_i \, \exp (-F[Q] ) \;, \quad
    F = F_0 + F_1 \;, \label{b12} \\ F_0 [Q] &=
    {\textstyle\frac{1}{2}} \sum_{i,j=1}^M C_{ij}\, \mathrm{STr}
    \, Q_i\, s\, Q_j s + n \sum_{i=1}^M \mathrm{STr}\, \ln Q_i
    - \mathrm{i} E \sum_{i=1}^M \mathrm{STr} \,
    Q_i\, s \;, \\ F_1 [Q] &= - \mathrm{i} \sum_{i=1}^M \mathrm{STr}\,
    Q_i\, s\,(\mathcal{E} - E \cdot \mathbf{1} + J_i) \;,
\end{split}
\end{equation}
where $E$ is the ``center of mass'' of the energy parameters in
$\mathcal{E}$. This is the most general supermatrix model for
ensembles of Gaussian random matrices with unitary symmetry
$\mathrm{U}(n)^M$ ($M = N/n$). Note that no approximations have been
made, and (\ref{b12}) is an \emph{exact} reformulation of the
original problem. The entries of the variance matrix $C_{ij}$ are
required to be real symmetric and positive, but are otherwise
arbitrary. The size of the supermatrices $Q$ depends on the
correlation function to be calculated and is the same as for the
supermatrices $Q$ of the standard non-linear sigma model
\cite{efetov83,efetov}. However, while the eigenvalues of the usual
sigma model field $Q$ are constrained to be $\pm 1$, the eigenvalues
of our superbosonization field $Q_i$ fluctuate; they are real and
positive in the boson-boson sector and unitary (i.e., of unit
modulus) in the fermion-fermion sector.

Let us also point out that, since the key formula (\ref{bosonize})
holds for a large class of functions $f$ (not just the Gaussian
functions), the present method is not restricted to Gaussian disorder
distributions. In the case of a more general disorder distribution,
the term $\exp \left( - {\textstyle\frac{1}{2}} \sum_{i,j=1}^{N}
C_{ij}\, \mathrm{STr} \, Q_i\, s\, Q_j s \right)$ in (\ref{b12}) is
replaced by a functional which is determined by the Fourier transform
of the disorder distribution function.

It should also be clear that such a description as (\ref{b12}) exists
even for $n < q\,$. We just need to replace the integration domain
and integration measure $DQ_i\, \mathrm{SDet}^n(Q_i)$ by the modified
one constructed in Section \ref{sect:KY}.

\section{Density of states for almost diagonal matrices}
\label{sect:almost}

We now demonstrate how the method developed in the previous section
works for the density of states of almost diagonal random Hermitian
matrices. We focus on the $n = 1$ case, which was investigated in
\cite{YevKr2004}, and we will compare the results of this reference
with ours. Note that the limit of almost diagonal random matrices is
not accessible (for $n = 1$) via the standard non-linear sigma model
for random matrix problems.

To calculate the density of states from the generating function
(\ref{b12}), we set $p = q = 1$ and $s = \mathbf{1}$. Our
superbosonization field $Q_i$ is now a $2 \times 2$ supermatrix
\begin{displaymath}
    Q_i = \begin{pmatrix} a_i &\sigma_i\\ \tau_i &b_i \end{pmatrix}
\end{displaymath}
with Berezin integral form
\begin{displaymath}
    DQ_i = (2\pi\mathrm{i})^{-1} \, db_i\, da_i\, \partial_{\tau_i}
    \partial_{\sigma_i} \;.
\end{displaymath}
The density of states per unit length for a system of $N$ sites is
expressed as
\begin{equation}\label{f1}
    \rho(E) = (2\pi N)^{-1} \, \mathfrak{Im}\,\sum_{i=1}^N
    \mathrm{i} \langle \mathrm{Tr}\, Q_i \rangle \;,
\end{equation}
where $\langle \ldots \rangle$ means the average with respect to the
statistical weight $\mathrm{e}^{-F[Q]}$ of (\ref{b12}), taken with
vanishing source term $J_i = 0\,$. The diagonal matrix of energy
parameters is $\mathcal{E} = \mathrm{diag}(E,E)$ with $\mathfrak{Im}
\, E > 0\,$.

We now proceed by first solving the diagonal variance matrix $C_{ij}
= C_0\, \delta_{ij}$ case exactly and then expanding in the
off-diagonal terms $C_{ij}\,$, $i\neq j$. For notational simplicity
we assume that the system is one-dimensional with
translation-invariant $C_{ij} = c(|i-j|)$ (although neither
assumption is really necessary).

In the zeroth order expansion the integral $\frac{1}{2}\langle
\mathrm{Tr}\, Q_i \rangle = \langle b_i \rangle = \langle a_i
\rangle$ in (\ref{f1}) factors as a product of $N$ independent
integrals, one for each site. The $N-1$ integrals for the sites $j
\not= i$ all are unity:
\begin{displaymath}
    \int DQ_j \, \mathrm{SDet}(Q_j)\, \mathrm{e}^{\mathrm{i} E
    \, \mathrm{STr}\, Q_j - (C_0/2)\, \mathrm{STr}\, Q_j Q_j} = 1
    \qquad (C_0 = C_{jj}) \;,
\end{displaymath}
which is a consequence of supersymmetry. The remaining integral for
the distinguished site $i$ is
\begin{eqnarray}
    &&\int DQ_i \, \mathrm{SDet}(Q_i)\, a_i \, \mathrm{e}^{\mathrm{i} E
    \, \mathrm{STr}\, Q_i - (C_0/2)\, \mathrm{STr}\, Q_i Q_i} \nonumber \\
    &=& (2\pi\mathrm{i})^{-1} \int_{\mathbb{R}_+} db \oint_{\mathrm{U}(1)}
    da \, \partial_\tau \partial_\sigma\, (b - \tau a^{-1} \sigma)\,
    \mathrm{e}^{\mathrm{i}E(b-a) - (C_0/2) (b^2 - a^2- 2\sigma\tau)} =
    \int_0^\infty \mathrm{e}^{\mathrm{i}E b - (C_0/2) b^2} db
    \label{eq:firstave}\;,
\end{eqnarray}
where in the second line we dropped the index $i$ from the
integration variables. The $\sigma \tau$ term in the exponent cannot
contribute to the Fermi integral $\partial_\tau \partial_\sigma\,$,
as one must pick the term $- \tau a^{-1} \sigma$ in front of the
exponential in order to have a non-zero integral over $a \in
\mathrm{U}(1)$. Hence, denoting the density of states in the zeroth
order of this expansion by $\rho_{d}(E)$, we have (for
$\mathfrak{Im}\, E \to 0+$)
\begin{equation}\label{f2}
    \rho_{d}(E) = \pi^{-1} \mathfrak{Im}\, \mathrm{i} \langle a_1 \rangle
    = (2\pi)^{-1} \int_{\mathbb{R}} \mathrm{e}^{\mathrm{i} E b - (C_0/2)
    b^2} db = (2\pi C_0)^{-1/2} \mathrm{e}^{- E^2 / 2C_0} \;.
\end{equation}
This Gaussian function is of course none other than the probability
distribution function of the diagonal elements of our almost diagonal
random matrix.

Now we calculate the correction $\delta \rho_{d}(E)$ coming from the
off-diagonal elements. For an almost diagonal variance matrix
$C_{ij}$ we may approximate the generating function $Z[J]$ by
expanding the exponential in the off-diagonal terms:
\begin{displaymath}
    Z[J] = \int \prod_{i=1}^N DQ_i \,\mathrm{SDet}(Q_i) \, \mathrm{e}^{
    \mathrm{i} \, \mathrm{STr} \, Q_i (\mathcal{E} + J_i) - (C_0/2) \,
    \mathrm{STr}\, Q_i Q_i} \Big(1 - {\textstyle\frac{1}{2}} \sum_{j
    \neq k} C_{jk}\,\mathrm{STr}\, Q_j Q_k + \ldots \Big) \;.
\end{displaymath}
Therefore the correction to $\rho_{d}(E)$ may be written as
\begin{equation}\label{f8}
    \delta\rho(E) = - (4\pi N)^{-1} \mathfrak{Im}
    \,\mathrm{i} \sum_{i} \sum_{j\neq k} C_{jk} \left\langle
    \mathrm{Tr} (Q_i) \,\mathrm{STr} (Q_j Q_k) \right\rangle \;.
\end{equation}
Again, $\langle \ldots \rangle$ is a product of $N$ independent
integrals, $N-3$ of which are unity by supersymmetry. Using the
relation $\langle Q^{\mu\nu} \rangle = \frac{1}{2} \langle
\mathrm{Tr}\, Q \rangle \, \delta^{\mu\nu}$ and $\mathrm{STr}\,
\mathbf{1} = 0\,$, we see that $\langle \ldots \rangle$ in (\ref{f8})
vanishes unless $i = j$ or $i = k\,$. If $i = k \neq j\,$ then (note:
2 angled brackets included in r.h.s.)
\begin{displaymath}
    \langle \mathrm{Tr}(Q_i) \,\mathrm{STr} (Q_j Q_i) \rangle
    = {\textstyle\frac{1}{2}} \langle \mathrm{Tr}(Q_i) \,
    \mathrm{STr}(Q_i)\rangle \, \langle\mathrm{Tr}(Q_j) \rangle \;.
\end{displaymath}
The single-site average of $\mathrm{Tr}\, Q_j$ is given by
(\ref{eq:firstave}). The expression for the single-site average of
$\mathrm{Tr}(Q_i) \, \mathrm{STr}(Q_i)$ is the same except that an
additional factor $\mathrm{STr}\, Q_i = b_i - a_i$ has to be inserted
under the integral sign. The term $-a_i$ of this factor gives
vanishing residue at the simple pole $a_i = 0$ and hence does not
contribute. So only an extra factor of $b_i \equiv b$ remains and we
get
\begin{displaymath}
    \int DQ_i \, \mathrm{SDet}(Q_i)\, \mathrm{Tr}(Q_i)\,
    \mathrm{STr}(Q_i) \, \mathrm{e}^{\mathrm{i} E \, \mathrm{STr}\,
    Q_i - (C_0/2)\, \mathrm{STr}\, Q_i Q_i} = \int_0^\infty
    \mathrm{e}^{\mathrm{i}E b - (C_0/2) b^2} b \, db \;.
\end{displaymath}
Altogether we then obtain for the correction to $\rho_d(E)$ (when the
variance matrix $C_{ij}$ is almost diagonal):
\begin{equation}
    \delta \rho(E) = - \frac{C_1}{2 C_0} \, \frac{d}{dE}\, \left(
    \mathrm{e}^{-E^2 / C_0} \mathrm{erfi}(E / \sqrt{2C_0}) \right)\;,
\end{equation}
where $C_1 = \sum_{j\neq 1} C_{1j}$ and $\mathrm{erfi}(x)$ is the
imaginary error function,
\begin{displaymath}
    \mathrm{erfi}(x) = \frac{2}{\sqrt{\pi}} \sum_{n=0}^\infty
    \frac{x^{2n+1}} {n! (2n+1)} \;.
\end{displaymath}
Note that $\delta\rho(E)$ is a total derivative. This is as expected
as turning on the off-diagonal elements of $C_{ij}$ neither creates
nor destroys levels but just changes their positions.

The high-energy limit of the density of states for almost diagonal
random matrices was calculated in Ref.\ \cite{YevKr2004}. To compare
with those results, we note that the large-$x$ limit of the imaginary
error function is $x\, \mathrm{e}^{-x^2} \mathrm{erfi}(x) \to
\sqrt{\pi}\,$, which gives
\begin{displaymath}
    \delta\rho(E) = \frac{C_1}{C_0} \rho_{d}(E) \qquad
    (E \gg \sqrt{2C_0}) \;.
\end{displaymath}
Thus we have the simple result that the density of states for an
almost diagonal Gaussian random matrix at high energies (i.e., in the
far tail of the Gaussian distribution) is
\begin{equation}
    \rho(E) = \sum_j (C_{ij}/C_{ii}) \, \rho_{d}(E) \qquad
    (E \gg \sqrt{2C_{ii}}) \;.
\end{equation}
This result is equivalent to that calculated in Ref.\
\cite{YevKr2004}.

\section{Reduction to the standard sigma model}
\label{sect:reduction}

Let us now see how the generalized supermatrix model (\ref{b12})
reduces to the standard diffusive non-linear sigma model under
suitable conditions. In order for such a reduction to take place, the
superbosonization field must get localized in a certain low-energy
submanifold (corresponding to the standard sigma model field) of the
total field space. As we shall see, the latter comes about if the
number of orbitals $n$ is large. If, in addition, the
superbosonization field has enough collectivity (or stiffness) due to
a variance matrix $C_{ij}$ with sufficiently long range, then the
effective degrees of freedom of the problem are the Goldstone modes
associated with the low-energy manifold, and one recovers the
diffusive non-linear sigma model.

To identify the low-energy manifold, we must first understand the
symmetries of the functional $F_0[Q]$ in (\ref{b12}). Beginning with
the boson-boson sector, let $g \in \mathrm{GL}_q (\mathbb{C})$ be any
invertible complex $q \times q$ matrix and consider the
transformation
\begin{displaymath}
    A_i \mapsto A_i \;, \quad B_i \mapsto g B_i g^\dagger \;, \quad
    \sigma_i \mapsto \sigma_i g^\dagger \;, \quad\tau_i \to g \tau_i \;.
\end{displaymath}
This transformation is a symmetry of the Berezin integral form $DQ_i$
of (\ref{eq:DEF-DQ}). It is a symmetry of $F_0 [Q]$ if
\begin{equation}\label{eq:psu}
    g^\dagger s_0 g = s_0 \;, \quad s_0 =
    \begin{pmatrix} \mathbf{1}_{q_+} &0 \\ 0 &- \mathbf{1}_{q_-}
    \end{pmatrix} \;,
\end{equation}
where $s_0$ is the boson-boson part of $s\,$. The condition
(\ref{eq:psu}) singles out a pseudo-unitary subgroup of
$\mathrm{GL}_q (\mathbb{C})$. This non-compact group is denoted by
$\mathrm{U}(q_+,q_-)$.

Turning to the fermion-fermion sector, let $(g,h) \in \mathrm{U}(p)
\times \mathrm{U}(p)$ and consider the transformation
\begin{displaymath}
    A_i \mapsto g A_i h^{-1} \;, \quad B_i \mapsto B_i \;, \quad
    \sigma_i \mapsto g \sigma_i \;, \quad \tau_i \to \tau_i h^{-1} \;.
\end{displaymath}
Again this is a symmetry of $DQ_i\,$. In order for it to be a
symmetry of $F_0[Q]$ we must impose the condition $g = h\,$, which
singles out the diagonal subgroup $\mathrm{U}(p) \subset
\mathrm{U}(p) \times \mathrm{U}(p)$.

The product of groups $\mathrm{U}(p) \times \mathrm{U}(q_+ , q_-)$ is
the group of bosonic symmetries of our problem. There also exist a
number of fermionic symmetries. It is not difficult to show that on
inclusion of these symmetries the symmetry group becomes a Lie
supergroup $\mathrm{U} (p\, | \, q_+,q_-)$. This is known to be the
symmetry group of the standard non-linear sigma model (for systems in
the unitary symmetry class). Note that the group action of
$\mathrm{U} (p\, | \, q_+ , q_-)$ is by conjugation $Q_i s \mapsto U
Q_i s\, U^{-1}$.

The next step is to look for minima of the 'energy' functional
$F_0[Q]$ in (\ref{b12}). (The term $F_1$ is considered small for
present purposes). Varying this functional gives the saddle-point
equation
\begin{equation}\label{eq:speq}
    0 = \frac{\delta}{\delta Q_i} F_0[Q] = \sum_j C_{ij} s\,
    Q_j s - (Q_i)^{-1} n - \mathrm{i} E s \;.
\end{equation}
Assuming translation invariance of $C_{ij}$ (and setting $C := C_0 +
C_1 = \sum_j C_{ij}$) one first looks for $i$-independent solutions
$Q_i \equiv Q$ in the space of diagonal matrices. From
(\ref{eq:speq}) each eigenvalue of the diagonal matrix $Q$ satisfies
a quadratic equation, which in general has two different solutions,
$q^{(+)}$ and $q^{(-)}$. In the boson-boson sector, one of these
solutions, say $q^{(-)}$, is ruled out by the positivity condition
$B_i > 0\,$. However, in the fermion-fermion sector both solutions
$q^{(\pm)}$ are in principle admissible. Connected supermanifolds of
solutions of the saddle-point equation are then generated by the
conjugation action $Q s \mapsto U Q s\, U^{-1}$ of the symmetry
group. The saddle-point manifolds thus obtained are orbits of
$\mathrm{U}(p \, | \, q_+ , q_-)$ which can be classified by the
number of eigenvalues $q^{(+)}$ and $q^{(-)}$ of the solution in the
fermion-fermion sector. For energies 'inside the band', $|E| <
\sqrt{4Cn}\,$, all these supermanifolds are parameterized by
\begin{equation}
    Q s = C^{-1/2} (n - E^2 / 4C)^{1/2} \, U \Lambda U^{-1} +
    \mathrm{i} (E/2C)\, \mathbf{1} \;, \qquad U \in
    \mathrm{U}(p\,|\, q_+ , q_-)\;,
\end{equation}
where $\Lambda$ is a diagonal matrix with eigenvalues $\pm 1:$
\begin{displaymath}
    \Lambda = \mathrm{diag}(s_1 , s_0) \;.
\end{displaymath}
The boson-boson part of $\Lambda$ is uniquely determined by the
boson-boson part $s_0$ of $s\,$, but the signs in the fermion-fermion
part $s_1$ of $\Lambda$ are arbitrary. In the general situation, each
of these saddle-point manifolds (corresponding to different choices
of $\mathrm{Tr}\, s_1$) make a contribution to the generating
function (\ref{b12}).

In the following we focus on the important case $p = q\,$. By making
the maximally supersymmetric choice $s_1 = s_0\,$, one obtains a
distinguished saddle-point manifold which dominates (under conditions
to be specified), while contributions from other saddle-point
manifolds are suppressed by fermionic zero modes due to a breaking of
supersymmetry. Low-energy configurations of the superbosonization
field are now expressed as
\begin{equation}\label{eq:sp-param}
    Q_i s = \pi\nu\, \tilde{Q}_i + \mathrm{i} (E/2C) \, \mathbf{1}
    \;, \quad \nu = \pi^{-1} C^{-1/2} (n - E^2 / 4C)^{1/2} \;,
\end{equation}
where $\tilde{Q}_i$ is a dimensionless field, namely the standard
sigma model field
\begin{displaymath}
    \tilde{Q}_i = U_i^{\vphantom{-1}}\, \Lambda\, U_i^{-1} \;, \quad
    \Lambda = \mathrm{diag}(s_0 , s_0) \;, \quad U_i \in \mathrm{U}
    (p \, | \, p_+ , p_-) \;.
\end{displaymath}

By computing the second variation of $F_0[Q]$ at the minimum, one
sees that the fluctuations of modes transverse to the low-energy
manifold (the so-called massive modes) are controlled by the
quadratic form
\begin{displaymath}
    h_{ij} = (\pi \nu)^2 C_{ij} + n\, \delta_{ij} \stackrel{E=0}{=}
    n (\delta_{ij} + C_{ij} / C)\;.
\end{displaymath}
(Here the second expression makes the simplifying assumption that $E
= 0\,$.) Since this quadratic form is bounded from below by the
diagonal form $n \, \delta_{ij}\,$, its eigenvalues are never smaller
than $n$ and it therefore follows that fluctuations of the massive
modes are strongly suppressed in the limit of $n \gg 1\,$.

In the case of small $n\,$, when $C_{ij}$ is short ranged (i.e., our
random matrix is almost diagonal), the massive modes fluctuate
strongly and there is no controlled reduction to the low-energy
manifold of fields $\tilde{Q}_i\,$. One might now think that the
situation gets better when $C_{ij}$ is taken to be long ranged.
However, this is definitely \emph{not} true when the
superbosonization method is used! The problem is that the Fourier
transform $\tilde{C}(k)$ of a long-ranged variance matrix $C_{ij} =
c(|i-j|)$ decreases with increasing wave number $k$ and is close to
zero at high wave numbers (near the edge of the Brillouin zone),
which implies that our high-momentum massive modes always have a
small mass (of order $n$) when $n$ is small.

It should be stressed that the case of a long-ranged variance matrix
is much better handled by the traditional Hubbard-Stratonovich
trans\-formation approach. There, fluctuations of the massive modes
(say at $E = 0$) are controlled by another quadratic form,
\begin{displaymath}
    h_{ij}^{(\mathrm{HS})} = n (\delta_{ij} + C w_{ij}) \;,
\end{displaymath}
where $w_{ij}$ is the matrix inverse of $C_{ij}$ (i.e., $\sum_j
w_{ij} C_{jl} = \delta_{il}$). The Fourier transforms of $C_{ij}$ and
$w_{ij}$ are reciprocals of each other, $\tilde{w}(k) = \tilde{C}
(k)^{-1}$. Thus, if the eigenvalues of $C_{ij}$ become small near the
edge of the Brillouin zone, then those of $w_{ij}$ become large, and
therefore the high-momentum massive modes of the Hubbard-Stratonovich
approach are truly massive for the case of long-ranged $C_{ij}$ (and
all $n$ including $n = 1$) and can be integrated out in a controlled
way. In summary, since a different quadratic form enters the game,
the reduction to the standard sigma model in the present approach is
valid under conditions that are not identical to those of the
Hubbard-Stratonovich approach.

We now insert (\ref{eq:sp-param}) into the expression (\ref{b12}) for
$F_0$ to obtain an effective energy functional for the $\tilde{Q}_i$
field:
\begin{displaymath}
    F_0 \approx {\textstyle{\frac{1}{2}}} \sum_{i,j} \tilde{C}_{ij}
    \, \mathrm{STr}\, \tilde{Q}_i \tilde{Q}_j \;, \quad \tilde{C}_{ij}
    = (\pi \nu)^2 C_{ij} \;.
\end{displaymath}
This is a good approximation when $n \gg 1\,$, in which case we may
simply neglect the massive modes. On the other hand, if $n$ is small
the massive modes fluctuate strongly and we face the non-trivial task
of integrating them out with a non-perturbative calculation. While
this will not change the symmetries of the effective energy
functional for the $\tilde{Q}_i$ field, it may cause a major
renormalization of the coupling parameters $\tilde{C}_{ij}\,$.

Therefore, to maintain quantitative control of the situation, we now
assume $n$ to be large. Taking into account the term $F_1$ in
(\ref{b12}), we arrive at a low-energy effective action
\begin{equation}\label{eq:LEEA}
    \tilde{F}[\tilde{Q}] = {\textstyle{\frac{1}{2}}}
    \sum_{i \not= j} \tilde{C}_{ij}\, \mathrm{STr}\,
    \tilde{Q}_i \tilde{Q}_j - \mathrm{i}\pi\nu \sum_{i=1}^M
    \mathrm{STr}\, \tilde Q_{i} (\mathcal{E} + J_i) \;,
\end{equation}
where diagonal terms have been dropped from the double sum over $i ,
j$ because $\mathrm{STr}\, \tilde{Q}_i \tilde{Q}_i = 0\,$. Recall
that $\tilde{Q}_i$ is the standard sigma model field.

The action (\ref{eq:LEEA}) describes a spectrum of physical
situations ranging from strong localization to diffusive behavior.
Strong localization occurs in the limit of almost diagonal random
matrices, where our coupling coefficients $\tilde{C}_{ij}$ $(i \not=
j$) are small, leading to a disordered sigma model field $\tilde{Q}_i
\,$. (Note that small $\tilde{C}_{ij}$ can be realized in spite of
$n$ being large.)

The diffusive regime can be realized by taking the variance matrix
$C_{ij}$ to be long ranged. In this limit, spatial variations of
$\tilde{Q}_i$ are suppressed and we may make a continuum
approximation $\tilde{Q}_i \to \tilde{Q}(r)$, expanding the first
term on the right-hand side of (\ref{eq:LEEA}) in gradients of
$\tilde{Q}(r)$. A standard computation gives the diffusive action
\begin{displaymath}
    {\textstyle{\frac{1}{2}}} \sum_{i \not= j} \tilde{C}_{ij}\,
    \mathrm{STr}\, \tilde{Q}_i \tilde{Q}_j \approx
    - {\textstyle\frac{\pi\nu}{8}}\,\mathrm{STr}\,\int dr\, D
    \big(\nabla \tilde{Q}(r) \big)^2 \;,
\end{displaymath}
where the `diffusion coefficient' is $D = 2\pi\nu \sum_j |i-j|^2
C_{ij}\,$. To summarize, the discussion above has shown how, in some
cases, the conventional non-linear sigma model can be obtained by
reduction from the supermatrix model (\ref{b12}).

Now in various regimes such a reduction cannot be done and one has to
work with (\ref{b12}). Among these are models with a critical point
where an Anderson metal-insulator transition takes place. Much
attention has been given to such models
\cite{evers,mirlin00,ndawana,YevKr2004} with the aim of better
understanding the critical behaviour at the transition. It is an
interesting open question whether the supermatrix model (\ref{b12})
can be of use in the analysis of these critical models.

\section{Conclusion}\label{sect:conclusion}

The main outcome of this paper is an exact mapping of a Gaussian
random matrix problem to a supermatrix model. The measure of
integration and the structure of the matrix field $i \mapsto Q_{i}$
are given in Eq.\ (\ref{eq:DEF-DQ}) for $n \geq q$ and in Eq.\
(\ref{eq:KY-formula}) for $n < q\,$. One might be tempted to call the
supermatrix model a `generalized non-linear sigma model', but such a
terminology would be misleading as the target space of the
supermatrix model is not homogeneous (unlike with what is called a
non-linear sigma model in the strict sense of the word) with respect
to its symmetry group.

A notable result is that the superbosonization formulas
(\ref{bosonize}) and (\ref{eq:KY-formula}) are {\it different},
depending on the value of the number of `orbitals' $n$ relative to
the number of commuting variables $q$ in the field $\psi\,$. The
former determines the \emph{real space} structure of the random
matrix ensemble, while the latter is equal to the number of points
$q$ in the $q$-point correlation function in the \emph{energy space}.
Thus the superbosonization formulas (\ref{bosonize}) and
(\ref{eq:KY-formula}) indicate that there exists a correlation, in a
certain sense, between real space and energy space. This fact seems
to be fundamental, although its implications are not yet fully
understood.

In this context, let us make one observation which we consider to be
of relevance. It has been a puzzle for many years now how an inspired
use of the replica trick with fermionic replicas correctly reproduces
\cite{KM1,KM2,YL,critique} the DoS-DoS correlation function in the
large-$n$ random matrix limit, whereas the replica trick with bosonic
replicas is known to fail \cite{VZ} when the same limit is invoked.
Based on the difference between $q \le n$ and $q > n$ observed above,
we propose the following resolution of this long-standing puzzle.

In the replica trick one needs to calculate the observable of
interest for every number of replicas and then, using all this
information, one tries to find an analytic continuation to zero
replica number. When using the replica trick with $q$ bosonic
replicas, we conjecture that vital information about the analytic
continuation $q \to 0$ is contained in the \emph{high} replica
numbers $q > n\,$, where the behavior of the correlation functions
seems to be qualitatively different (based on what we have seen with
the superbosonization formula) from that for $q \le n\,$. From this
vantage point, it has to be regarded as an ill-advised scheme to take
the large-$n$ limit (and thus the saddle-point approximation leading
to the bosonic replica NL$\sigma$M) at an early stage of the
calculation, as was done in \cite{VZ}. Indeed, in the process of
taking $n \to \infty$ the $q > n$ branch, and hence all information
carried by it, is lost from the computation. Our explanation of the
long-standing puzzle is that it is the loss of the $q > n$ branch
which makes it impossible to construct the correct analytic
continuation to zero replica number.

This scenario and other questions related to the difference between
the cases $n \geq q$ and $n < q$ are interesting directions for
future research. Here, in contrast, we outline the idea of a
\emph{uniform} representation of both cases which requires a
non-trivial limiting procedure. For simplicity we restrict ourselves
to the case of $n = 1$, $p = q = 2\,$. As was explained in detail at
the beginning of Sect.\ \ref{sect:KY}, the simple superbosonization
formula (\ref{bosonize}) fails in this case. For the sake of
discussion let us recall the details here. The argument is especially
transparent when the function $F(Q)$ depends only on the commuting
variables in the matrices $A$ and $B$ and is non-zero in the limit $B
\rightarrow 0$. Note that the left-hand side of (\ref{bosonize}) is
non-zero in general. On the other hand, since $A$, $B$, $\sigma$, and
$\tau$ for $p = q = 2$ are $2\times 2$ matrices, the superdeterminant
\begin{displaymath}
    \mathrm{SDet}(Q) = \mathrm{Det}(B) \, \mathrm{Det}^{-1}(A)
    \, \mathrm{Det}(1 - A^{-1} \sigma B^{-1} \tau)
\end{displaymath}
is a quartic polynomial in eight Grassmann variables
$\sigma^{\mu\nu}$ and $\tau^{\nu\mu}$ ($\mu,\nu = 1, 2$) and
therefore $\prod \partial_{\tau^{\nu\mu}} \partial_{\sigma^{\mu\nu}}
\, {\rm SDet}(Q) = 0\,$. This means that the right-hand side of
(\ref{bosonize}) is zero, unless there is a divergence in the
integration over commuting variables. Here one must pay attention to
the fact that the integral over the bosonic variables $B$ can be
singular along the $\mathrm{Det}(B) = 0$ boundary. A quick way to see
this is to make a transformation to new variables $\sigma' = A^{-1}
\sigma$ and $\tau' = B^{-1} \tau\,$, which has Jacobian
$\mathrm{Det}^{-q} (A)\, \mathrm{Det}^{-q} (B)$ and leads to a
singular dependence $\mathrm{Det}^{1-q}(B) = \mathrm{Det}^{-1}(B)$ at
$q = 2\,$. Thus, a naive application of the superbosonization formula
(\ref{bosonize}) to the case $n = 1$, $p = q = 2$ leads to an
ill-defined expression of the type $0 \times \infty\,$.

The idea of a unified description stems from the observation that the
deficiency of the Grassmann variables in $\mathrm{SDet}(Q)$ (where 4
are missing from the full set of 8), which is responsible for the $0$
part in $0 \times \infty$, does not persist when $\mathrm{SDet}(Q)$
is replaced by $\mathrm{SDet}^{1+\alpha}(Q)$ with $\alpha \in
\mathbb{C}$ being a small regularization parameter. Of course, to
move $\alpha$ off zero, we must first define the factor
$\mathrm{Det}^{-1-\alpha}(A)$ in $\mathrm{SDet}^{1 + \alpha}(Q)$.
This can be done by making some choice of fundamental domain for the
logarithm $\ln : \, \mathrm{GL}_2(\mathbb{C}) \to \mathbb{C}\,$.
Restricting $A \in \mathrm{U}(2) \subset \mathrm{GL}_2(\mathbb{C})$
to this fundamental domain, we may expand
\begin{displaymath}
    \mathrm{SDet}^{1+\alpha}(Q) = \mathrm{SDet}(Q) \, \left(1 + \alpha \,
    \ln \mathrm{SDet}(Q) + ... \right) \;.
\end{displaymath}
One then easily verifies that the Fermi integral of $\mathrm{SDet}^{
1+\alpha}(Q)$ is non-zero for small but non-zero $\alpha\,$. At the
same time, the singular factor $\mathrm{Det}^{-1} (B)$ in the
integrand is replaced by $\mathrm{Det}^{\alpha-1}(B)$, which makes
the singularity at $B\rightarrow 0$ integrable for $\mathfrak{Re}\,
\alpha > 0\,$. Moreover, it is possible to analytically continue the
integral over $A \in \mathrm{U}(2)$ to $\alpha \not=0$ by taking the
integration contour, i.e., the 4-dimensional real submanifold of
integration in $\mathrm{GL}_2(\mathbb{C})$, to infinity along the
boundary of the fundamental domain of the logarithm. (This will work
if the boundary of the fundamental domain has been chosen as a
submanifold of rapid decay of the integrand.) Thus, carefully taking
the limit $\lim_{\alpha \to 0+} \mathrm{SDet}^{1 + \alpha}(Q)$
instead of immediately setting $\alpha = 0\,$, is a way to give
meaning to the ill-defined expression $0 \times \infty\,$.

Exploratory calculations done along these lines give a remarkable
coincidence with Eq.\ (\ref{eq:KY-formula}). Our explorations suggest
that the superbosonization formula in both the $n = q = 1$ and $n =
1$, $q = 2$ cases can be written in a uniform way resembling the
original form of Ref.\ \cite{est} and Eq.\ (\ref{bosonize}) but with
$\lim_{\alpha \rightarrow 0+} \mathrm{SDet}^{1 + \alpha}(Q)$
replacing $\mathrm{SDet}(Q)$. It should be stressed, however, that
the mathematics at hand gets more complicated as the number of
replicas $p = q$ goes up. We have not yet made a serious effort to
confront these complications and more work is needed to put the idea
of $\alpha$-regularization on a solid mathematical basis in the
general case.

To finish this Conclusion, we would like to compare our
superbosonization (SB) approach with the standard one
\cite{efetov83,efetov} based on the Hubbard-Stratonovich (HS)
transformation and the degenerate saddle-point approximation. As was
mentioned in the Introduction, the main disadvantage of the standard
approach is that it does not apply to random matrix ensembles with a
short-ranged variance matrix $C_{ij}\,$. Another disadvantage is that
it is applicable only to Gaussian ensembles. Both restrictions are
lifted in the new SB approach. The new approach is in some sense dual
to the standard HS approach as it relies on the variance matrix
$C_{ij}\,$, while the HS approach relies on its \emph{inverse},
$(C^{-1})_{ij}\,$. This feature makes the new SB approach most
efficient for short-ranged $C_{ij}\,$, e.g., for almost diagonal
random matrices \cite{KrYev2003}. It can also be used for long-ranged
$C_{ij}$ but only under the condition that the number of orbitals $n$
is large. For $n = 1$, extracting the standard NL$\sigma$M from Eqs.\
(\ref{b12}) is a highly non-trivial task for the case of a
long-ranged matrix $C_{ij}$ (which is a textbook example of the
derivation of NL$\sigma$M in the framework of the HS approach).

This example shows once again that the two exact supermatrix
representations of the random matrix ensemble which emerge after the
SB and HS transformations, are fundamentally different and largely
complementary. Thus the new SB approach does not negate or supersede
the standard HS one. Rather, it extends the supersymmetry method to
an area previously not accessible to it.

K.B.E.\ and M.R.Z.\ acknowledge financial support by the Deutsche
Forschungsgemeinschaft (SFB/TR 12).

\end{document}